\documentclass[sigconf]{acmart}

\AtBeginDocument{%
  }

\usepackage{graphicx}
\graphicspath{ {./img/} }

\usepackage{amsmath}
\usepackage{amssymb}
\usepackage[ruled,vlined]{algorithm2e}
\usepackage{subcaption}
\usepackage{tikz}
\usepackage{enumitem}
\usepackage{textcomp}
\usepackage{listings}
\usepackage{comment}
\usepackage{balance}
\usepackage{gensymb}
\usepackage{hyperref}
\usepackage{cleveref}

\setcopyright{acmlicensed}
\copyrightyear{2024}
\acmYear{2024}
\acmDOI{10.1145/3678717.3691320}

\acmConference[SIGSPATIAL '24]{The 32nd ACM International Conference on Advances in Geographic Information Systems}{October 29-November 1, 2024}{Atlanta, GA, USA}
\acmBooktitle{The 32nd ACM International Conference on Advances in Geographic Information Systems (SIGSPATIAL '24), October 29-November 1, 2024, Atlanta, GA, USA}
\acmISBN{979-8-4007-1107-7/24/10}

\begin{document}

\title{Multi-Entry Generalized Search Trees for Indexing Trajectories}

\author{Maxime Schoemans}
\affiliation{%
  \institution{Universit\'e libre de Bruxelles}
  \country{}
}
\email{maxime.schoemans@ulb.be}

\author{Walid G. Aref}
\affiliation{%
  \institution{Purdue University}
  \country{}
}
\email{aref@purdue.edu}

\author{Esteban Zim\'anyi}
\affiliation{%
  \institution{Universit\'e libre de Bruxelles}
  \country{}
}
\email{esteban.zimanyi@ulb.be}

\author{Mahmoud Sakr}
\affiliation{%
  \institution{Universit\'e libre de Bruxelles}
  \country{}
}
\email{mahmoud.sakr@ulb.be}


\begin{abstract}
The idea of generalized indices is one of the success stories of database systems research. It has found its way to implementation in common database systems. GiST (Generalized Search Tree) and SP-GiST (Space-Partitioned Generalized Search Tree) are two widely-used generalized indices that are typically used for multidimensional data, e.g., to index spatial or spatio-temporal data. Currently, the  generalized indices GiST and SP-GiST represent one database object using one index entry, e.g., a scalar value or a bounding box for each spatial or spatio-temporal object. However, when dealing with complex objects, e.g., moving object trajectories, a single entry per object is inadequate for creating efficient indices. Previous research has highlighted that splitting trajectories into multiple sub-trajectories or bounding boxes prior to indexing can enhance query performance as it leads to a higher index filter. 
In this paper, we introduce MGiST and MSP-GiST, the multi-entry generalized search tree counterparts of GiST and SP-GiST, respectively, that are designed to enable the partitioning of objects into multiple entries during insertion. 
The methods for decomposing a complex object into multiple sub-objects differ from one data type to another, and may depend on some domain-specific parameters. Thus, MGiST and MSP-GiST are designed to allow for pluggable modules that aid in optimizing the split of an object into multiple sub-objects. 
We demonstrate the usefulness of MGiST and MSP-GiST using a trajectory indexing scenario, where we realize several trajectory indexes using MGiST and MSP-GiST and  instantiate these search trees with trajectory-specific splitting algorithms. We  create and test the performance of several multi-entry versions of widely-used spatial index structures, e.g., R-Tree, Quad-Tree, and KD-Tree. We conduct evaluations using both synthetic and real-world data, and observe up to an order of magnitude enhancement in performance of point, range, and nearest neighbor queries.
\end{abstract}





\received{07 June 2024}
\received[accepted]{23 August 2024}

\maketitle

\section{Introduction}
\label{section:introduction}

GiST~\cite{gist} and SP-GiST~\cite{spgist} are software engineering solutions that facilitate the creation of custom-tailored indices inside a database management system. 
They allow users to easily implement a new index by specifying a set of interface parameters and external methods specific to that index. Using these parameters and methods, GiST or SP-GiST will take care of the construction, storage, and search, as well as the concurrency and crash recovery of the index. For example, the GiST framework can be used to implement most balanced search trees, e.g., the B+-Tree, the R-Tree, and the RD-Tree using minimal implementation effort. On the other hand, SP-GiST can be used to construct space-partitioning trees, e.g., the Quad-Tree, the KD-Tree, and the Trie, and their variants. 
In PostgreSQL, these frameworks allow database extensions, e.g., PostGIS~\cite{postgis} or MobilityDB~\cite{mobilitydb} to easily construct R-Trees, KD-Trees, and Quad-Trees for their spatial and spatio-temporal data.

\begin{table}[ht]
    \centering
    \caption{Classes of generalized indices}
    \begin{tabular}{lccc}
    \toprule
    Single-Entry & B-Tree & GiST & SP-GiST \\
    Multi-Entry & GIN & \textcolor{red}{MGiST} & \textcolor{red}{MSP-GiST} \\
    \bottomrule
    \end{tabular}
    \label{tab:generalized_indices}
\end{table}

However, in their current implementation, GiST and SP-GiST store each tuple as a single entry in the index. When handling complex or composite data types, much information is lost when compressing the object into a single index entry. This  reduces the capabilities and efficiency of the constructed indices. The idea of storing multiple entries for a single tuple is already being used for one-dimensional data, e.g., as in the case of the text stream data type. 
The Generalized Inverted Index, GIN~\cite{gin}, indexes complex one-dimensional data types, e.g., text documents that are composed of simpler elements that one may want to search for, e.g., words and trigrams. While useful, GIN  indexes only one-dimensional objects that exhibit total order, i.e., are sortable.  Internally, and building on the total order property, GIN stores its entries in a B-Tree, but is not applicable to the multi-dimensional case.

Consider a set, say $S$, of elements that are stored inside the index, and a set, say $R$, of indexable real-world complex objects. One can view the GiST and SP-GiST indices as forming a 1-1 relationship between sets $S$ and $R$. In contrast, GIN provides an m-1 relationship between $S$ and $R$, e.g., that multiple words in the index correspond to one document object in the document database. Thus, the current GiST and SP-GiST can be viewed as special cases of the more general case, and they are actually 1-1 GiST and 1-1 SP-GiST. 
This paper highlights the importance of realizing the more general case of the m-1 GiST and m-1 SP-GiST indices, provides extensible designs for both indices, and study their performance.

This paper presents the MGiST (Multi-Entry GiST) and MSP-GiST (Multi-Entry SP-GiST) frameworks as multi-entry m-1 generalizations of the 1-1 GiST and the 1-1 SP-GiST frameworks. Through the use of one additional external method, one can specify a splitting and decomposing mechanism that the framework will apply on each real-world object (the one-side of the m-1 relationship) before the resulting sub-objects (the m-side of the m-1 relationship) get inserted inside the index.
All the $m$ index entries store the identifier of the object (the object's oid, for short) to point back from the index side to the same object/tuple in the relational table side. This allows for the creation of multi-entry variants of  known indices, e.g., R-Trees, Quad-Trees, and KD-Trees. These multi-entry indices can be queried in a manner similar to the traditional GiST and SP-GiST indices.
By exposing the object splitting mechanism as a pluggable module,  MGiST and MSP-GiST  can be tailored for many different complex and composite data types with minimal implementation effort from the user. Examples of the data types that can benefit from multi-entry m-1 indices are: complex geometry types, e.g., paths  and non-convex polygons;  geometry collections, e.g.,  multipoints or multipolygons; and trajectory data. Without loss of generality, in this paper, we focus on the case of trajectories. 

\begin{figure}
    \centering
    \begin{subfigure}{.46\linewidth}
        \centering
        \includegraphics[width=\linewidth]{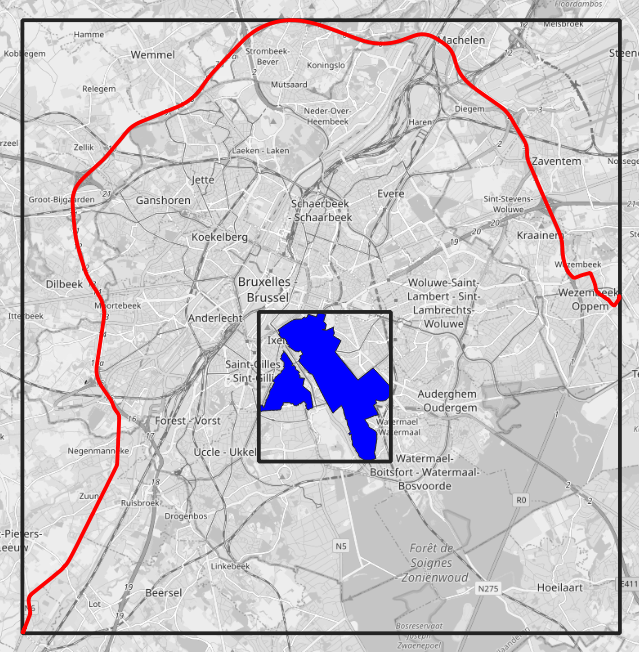}
        \caption{}
        \label{fig:intro_1}
    \end{subfigure}
    \begin{subfigure}{.48\linewidth}
        \centering
        \includegraphics[width=\linewidth]{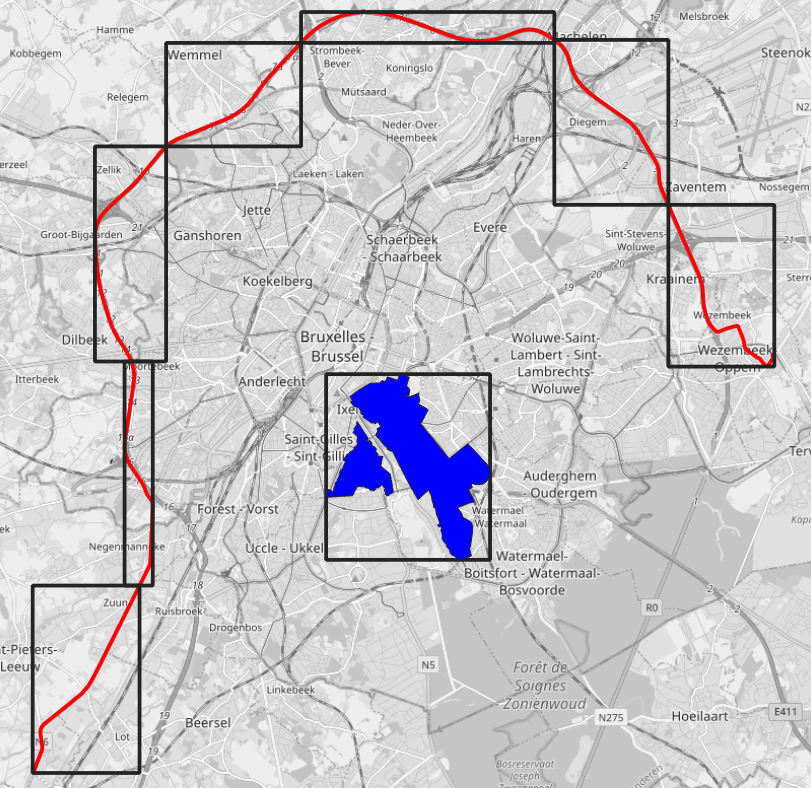}
        \caption{}
        \label{fig:intro_2}
    \end{subfigure}
    \caption{Representing a trajectory using multiple bounding boxes can reduce false positives of spatio-temporal indices.}
    \label{fig:intro}
\end{figure}

Trajectory data represents the movement of objects 
in space, e.g., vehicles or humans. This data can get large in two main aspects: the number of trajectories in a database, and the size of the individual trajectories. When observing the movement of objects over a long duration of time, the size of an individual trajectory can get large in terms of the number of observations, duration, and distance covered. The purpose of moving-object databases is to store and analyze these large trajectory data sets.

When analyzing large data sets, queries are commonly applied to subsets of the data. For example, one might be interested in computing the average speed of all cars passing through a given street. Alternatively, one might want to know which 10 cars have passed closest to a given location. To speedup these queries, databases use indices to filter out a large part of the data. In the case of trajectories, this is achieved by computing the spatio-temporal extent of a trajectory, and then indexing the resulting 3D or 4D bounding boxes (2/3D Spatial + 1D Temporal extent). This method works relatively well for short trajectories (i.e., small-sized trajectories), but fails when the trajectories are long because the bounding box does not accurately represent the trajectory anymore. Refer to Figure~\ref{fig:intro} for illustration, where the bounding box overlap (as tested by the index) would fail to filter out a trajectory that is clearly not passing through the region of interest. When working with long trajectories, these situations are common, and significantly impact the performance of point, range, and nearest neighbour queries.

MGiST and MSP-GiST  store a more detailed representation of the actual trajectory in the index. A trajectory is represented by a collection of smaller sub-trajectories that can each be indexed individually in MGiST and MSP-GiST. One extreme is to index each trajectory segment (pair of subsequent observations) individually. This method produces the best accuracy (in terms of false positives) but creates indices that are larger than the actual data, which is not desirable due to their large storage cost and long querying time. As a trade-off, one could split a trajectory into a set of smaller sub-trajectories, and index the bounding boxes of these sub-trajectories. Splitting the trajectory into a set of smaller bounding boxes produces a better approximation of the actual trajectory than representing it with a single bounding box, which  improves index efficiency by reducing the number of false positives (Fig.~\ref{fig:intro}). By limiting the number of bounding boxes, one can bound the total size of the index in case of storage constraints.

In this paper and as a proof of concept, we  instantiate a multi-entry R-Tree using MGiST, a multi-entry Quad-Tree and a multi-entry KD-Tree using MSP-GiST using several trajectory splitting strategies. We evaluate the performance of these instantiated indices for each of the splitting strategies, and demonstrate that multi-entry indices can improve point, range and nearest neighbor query performance by up to an order of magnitude when compared to their traditional single-entry counterparts.

The rest of this paper proceeds as follows. Sections~\ref{section:related_work} and~\ref{section:preliminaries} present related work, and overviews the GiST and SP-GiST frameworks. Section~\ref{section:mgist} presents the MGiST and MSP-GiST frameworks and describes how they can realize instances of multi-entry indices. As a case study, Section~\ref{section:traj_index} presents the implementation of multi-entry R-Trees, Quad-Trees, and KD-Trees for trajectories, together with their trajectory splitting algorithms. Section~\ref{section:experiments} presents the experimental evaluation of MGiST- and MSP-GiST-based indices using synthetic and real-world data sets. Finally, Section~\ref{section:discussion} discusses possible future work, and Section~\ref{section:conclusion}  concludes the paper.

The code implementing MGiST, MSP-GiST and their related instantiations for trajectory indexing is available online at \url{www.github.com/MobilityDB/mest}.

\section{Related Work}
\label{section:related_work}

\paragraph{GiST} To simplify the implementation of new indices, Hellerstein et al.~\cite{gist} introduce GiST, a generalized  search tree that is easily extensible for different data types and queries. GiST can instantiate balanced search trees, e.g., the B-Tree, R-Tree or M-Tree, using minimal implementation effort through the use of pluggable modules. Behind the scenes, GiST transparently handles the complex index internals, e.g.,  tree balancing, concurrency, and recovery~\cite{gist_concurrency}.

\paragraph{SP-GiST} With the emergence of new database applications, new indexing solutions, e.g., KD-Trees, Quad-Trees, Tries, and other unbalanced search trees have become more prominent. However, these indices could not be implemented in GiST, as they are not balanced trees. Aref et al.~\cite{spgist} introduce SP-GiST, a space-partitioning generalized index, that can instantiate space-partitioning unbalanced trees in a generalized framework. Many improvements and optimizations have been proposed for both  GiST and SP-GiST, e.g., ~\cite{gist_search,spgist_bulk,gist_scan} as well as their implementation in existing database systems, e.g.,~\cite{spgist_realization, rtree_penalty}.  GiST and SP-GiST are further detailed in Section~\ref{section:preliminaries}.

\paragraph{The Temporal Dimension} Many specialized indices have also been developed for the indexing of spatial and spatio-temporal data. The R-Tree~\cite{rtree}  indexes spatial objects  using their bounding boxes in a balanced tree structure. By storing the time as an additional dimension, spatio-temporal objects can be indexed in 3D R-Trees~\cite{3drtree} or RT-Trees~\cite{rttree_mrtree}. Another approach to index the temporal dimension is to have an R-Tree for each time instant. To save space, consecutive R-Trees can also share sub-trees that remain unchanged between two time instants, e.g., as in~\cite{rttree_mrtree, hrtree, hrplustree}. There are also  combinations of the above approaches, e.g., as in~\cite{rttree_mrtree}.
%the MR-Tree~\cite{rttree_mrtree}, HR-Tree~\cite{hrtree} and HR+-Tree~\cite{hrplustree}. 
%The MV3R-Tree~\cite{mv3rtree} combines the best of both approaches by grouping an MR-Tree and a 3D R-Tree into a single index structure.

\paragraph{Indexing Trajectories} A trajectory is a special type of spatio-temporal data. It combines multiple trajectory segments into a single large data object, with each segment being formed by linear interpolation of two stored instants. In the case of a primary index, each segment needs to be stored in the index, e.g., in a 3D R-Tree. STR-Trees and TB-Trees~\cite{strtree_tbtree} are extensions of the 3D R-Tree that cluster segments of the same trajectory to improve trajectory-related queries. 
If the tree is used as a secondary index, not all the segments have to be stored individually. Thus, one can index fewer bounding boxes that each represents a subset of the complete trajectory. This is the scenario handled by this paper. Hadjieleftheriou et al.~\cite{mergesplit} and Rasetic et al.~\cite{linearsplit}  present possible splitting algorithms for this purpose. These algorithms are re-used in the implementation of MGiST and MSP-GiST for trajectories as in Section~\ref{section:traj_index}. 

Another class of trajectory indices employs a 2-level approach. The first level partitions the space using a space-partitioning index, e.g., a grid or Quad-Tree. This index will be slowly changing, as new updates commonly appear close to existing values. In a second level, each spatial cell/partition contains a temporal index that stores the temporal information of the trajectory segments covered by this spatial partition. This approach is used in~\cite{seti,sebtree,csetree,trajstore}, with different combinations of spatial and temporal indices.

Most of the indices discussed above are mainly designed for range queries, with some being also able to answer nearest neighbour queries. Other trajectory indices have been presented to answer different types of queries. For example, the MTSB-Tree~\cite{mtsbtree}  can answer close pair queries that search for trajectories that have been closer than a given distance during a time interval and inside a given spatial range. The TrajTree~\cite{trajtree} index is used to match/query similar trajectories using the Edit Distance with Projections (EDwP) measure, and uses a structure similar to the R-Tree. Chebyshev polynomials~\cite{chebyshev} approximate and index multi-dimensional trajectories for the purpose of similarity matching. The PA-Tree~\cite{patree} is another index that uses Chebyshev polynomials to approximate spatio-temporal trajectories.
Lastly, many indices have been proposed for the purpose of indexing network-constrained trajectories as well as indoor trajectories. We do not detail these solutions, as they refer to a different problem. Refer to~\cite{survey_1,survey_2,survey_3} for  surveys on spatio-temporal access methods.
\section{Preliminaries: GiST and SP-GiST}
\label{section:preliminaries}

\textit{Generalized Search Tree} (GiST, for short)~\cite{gist} is a highly configurable generalized index structure that can instantiate balanced search trees supporting an extensible set of queries and data types. By specifying as little as six key methods, the GiST index can be used as, e.g., a B+-Tree, an R-Tree, or an RD-Tree.

\subsection{GiST Key Methods}
One needs to specify six key methods to instantiate a GiST index. These methods each serves a different purpose.

\textbf{Consistent($E$, $q$)}:  Method \textbf{Consistent} is used in the search algorithm of GiST.  Given an index entry $E = (p, ptr)$, it determines if the entry predicate $p$ is consistent with the query predicate $q$. This method only returns false if it is guaranteed that $p \wedge q$ is unsatisfiable.

\textbf{Compress($E$)} and \textbf{Decompress($E'$)}: These methods are used to store compressed representations of the entry predicates. Given an entry $E = (p, ptr)$, \textbf{Compress($E$)} returns a new entry $E' = (p', ptr)$, whose predicate $p'$ is a compressed representation of $p$. In contrast, given an entry $E' = (p', ptr)$, formed by compressing entry $E = (p, ptr)$, \textbf{Decompress($E'$)} returns a new entry $E'' = (p'', ptr)$, such that $p \rightarrow p''$. The notation $p \rightarrow p'$ implies that if predicate $p$ holds then $p'$ also holds. In the R-Tree, {\bf Compress} returns a bounding box given a complex spatial object, e.g., a polygon. {\bf Decompress} is the identity and is thus lossy as it is not possible to recover the initial geometry from its bounding box.

The remaining three methods are used during tree construction or when inserting a new tuple into the index.

\textbf{Union($P$)}: Given a set $P = \{E_1, ..., E_n\}$ of $n$ index entries $E_i = (p_i, ptr_i)$, determines a predicate $p'$ that holds for all entries in the set ($p_1 \vee ... \vee p_n \rightarrow p'$).

\textbf{Penalty($E_1$, $E_2$)}: Given two entries $E_1$ and $E_2$, computes a penalty value for inserting entry $E_2$ into the sub-tree rooted at $E_1$.

\textbf{PickSplit($P$)}: Given a set $P = \{E_1, ..., E_{M+1}\}$ of $M+1$ entries, splits $P$ into two subsets $P_1$ and $P_2$, each containing at least $k \cdot M$ entries, where $M$ is the maximum node capacity of the tree and $k$ corresponds to the minimum fill factor.

An example of  {\bf Union}  for spatial data  computes the minimum bounding box enclosing all the bounding boxes of the set of entries $P$.  {\bf Penalty}  would  compute a value related to the increase in area of the bounding box of $E_1$ caused by the insertion of $E_2$ into the sub-tree rooted at $E_1$. Lastly, a possible PickSplit method would split the set $P$ of entries by trying to minimize either the total area of, or the overlap between, the bounding boxes of $P_1$ and $P_2$.

\textbf{Distance($E_1$, $E_2$)}: 
GiST as realized in PostgreSQL has a seventh optional key method, \textbf{Distance}, that is used in answering nearest neighbor queries.
Given two entries $E_1$ and $E_2$, \textbf{Distance} computes a distance value between the predicates of $E_1$ and $E_2$.

\subsection{SP-GiST Key Methods and Parameters}

Similar to GiST, SP-GiST presents a list of key methods that are to be implemented to instantiate a specific space partitioning index. Methods \textbf{Consistent} and \textbf{PickSplit} are similar to those of GiST.

\textbf{Consistent($E$,$q$,$l$)}: Given Index Entry $E=(p,ptr)$ at Level $l$, determines if Predicate $E.p$ is consistent with Query Predicate $q$.

\textbf{PickSplit($P$, $l$)}: Given a set $P = \{E_1, ..., E_{M+1}\}$ of $M+1$ entries at Tree Level $l$, splits $P$ into $N$ partitions, where $M$ (maximum bucket size) and $N$ (number of partitions) are user-defined parameters.

SP-GiST can apply different partitioning methods based on the level. Thus, Tree Level $l$ is  passed as parameter to Consistent and PickSplit, e.g., the 2D KD-Tree partitions on the X-axis at even levels and on the Y-axis at odd levels.
SP-GiST  also has a \textbf{Cluster} method that clusters nodes into data pages. Although {\bf Cluster}  is described as a key method in~\cite{spgist}, it is implemented using a default clustering algorithm in PostgreSQL. 
Lastly,  SP-GiST supports nearest neighbor queries using a user-defined \textbf{Distance} method.
\section{Multi-Entry Search Trees}
\label{section:mgist}

A traditional index holds entries of the form: $E = (p, ptr)$, where $p$ is a predicate that holds for a given object and is used to search the tree, while $ptr$ is a pointer to said object. For example, in a PostgreSQL's R-Tree, $p$ is a bounding box and $ptr$ is the tuple ID ($\textit{TID}$) that references the physical position of the tuple on disk. Commonly, each $ptr$ only appears once in the index. That is, each tuple is indexed using at most one entry. 
%This is not the case for GIN (an inverted index) (as will be discussed further in Section~\ref{section:gin}).

In this section, we introduce MGiST and MSP-GiST, multi-entry generalized search trees that allow objects to be split into multiple entries before insertion. Each entry contains a different predicate but references the same tuple, and thus has the same $ptr$ value. This allows the predicates to be more restrictive, and can improve index efficiency by decreasing the number of matches to a given query predicate. For example, representing one trajectory using a set of small bounding boxes can reduce the number of false predicate matches (See Figure~\ref{fig:intro}). The application of multi-entry generalized search trees for indexing trajectories is discussed in Section~\ref{section:traj_index}.

MGiST and MSP-GiST contain all the key methods of GiST and SP-GiST as well as the additional method {\bf ExtractValue}. This method splits each tuple into a set of index entries (detailed in Section~\ref{section:extractvalue}). This addition changes both the insertion and search methods. 
%The changes to the insertion methods are minimal, while the changes to the search method are more significant. Indeed, w
With multiple entries pointing to the same tuple, de-duplication mechanisms are necessary when querying the index. Query semantics are also different from the ones in a traditional single-entry index. These changes are detailed below.

\subsection{ExtractValue Method}
\label{section:extractvalue}

{\bf ExtractValue} is a key method that splits an object into multiple index entries before insertion. After splitting, all returned entries are  inserted into the index using the existing GiST and SP-GiST insertion methods. Notice that nothing prevents  {\bf ExtractValue}  from returning only a single entry. However, MGiST and MSP-GiST   assumes that at least one tuple is split into more than one index entry. If all tuples are split into only 1 entry each, the traditional GiST and SP-GiST indices can be used.

\textbf{ExtractValue(E)}: Given an entry $E = (p, ptr)$, returns a set $P = \{E_1, ..., E_n\}$ of $n$ index entries, with $n \ge 1$, $E_i = (p_i, ptr)$.

{\bf ExtractValue} offers lot of flexibility. By deciding how the tuples are split and in how many parts, the created index can be tailored to specific use cases. Section~\ref{section:traj_index} details the particular use-case of trajectory indexing, and presents multiple different splitting algorithms that can be used with different purposes and effectiveness.

\subsection{Multi-Entry Search}
\label{section:search}

Searching an MGIST or MSP-GiST index consists of finding all entries that match the given query predicate. Each entry will  contain a pointer to a tuple that is part of the result of the query. 
%When working with multi-entry search trees, i
It is possible that multiple entries pointing to the same tuple match the query predicate. Thus, it is  necessary to apply a de-duplication step before returning the actual tuples. 
%Previous work~\cite{deduplication_1,deduplication_2} presents efficient techniques for tuple de-duplication in Quad-Tree-like structures. However, these methods only work when the entry predicate is duplicated. In this case, no assumption is made on the relation between different entry predicates of the same tuples. 
As no assumption is made on the relation between different entry predicates of the same tuples, existing de-duplication mechanisms~\cite{deduplication_1,deduplication_2} cannot be used.
Thus, duplicates are eliminated using a hashmap on tuple pointers ($ptr$). Before returning a tuple, the search algorithm checks that the pointer is not in the hashmap before returning the tuple. If the pointer is already in the hashmap, the tuple is skipped as it has already been returned before. Else, the tuple is returned and its pointer is added to the hashmap. 
%The M(SP-)GiST search algorithm is detailed in Algorithm~\ref{alg:search}.
% \begin{algorithm}
% \caption{M(SP-)GiST Search Algorithm}
% \label{alg:search}
% \KwIn{An M(SP-)GiST index rooted at $R$ and a query predicate $q$.}
% \KwOut{All the tuples matching $q$.}
% \If{MGiST}{
%     $P = $ GiST $Search(R, q)$\;
% }
% \ElseIf{MSP-GiST}{
%     $P = $ SP-GiST $Search(R, q)$\;
% }
% $T$ \tcc*{Empty list of tuple pointers}
% $M$ \tcc*{Empty hashmap}
% \For{Entry $E_i = (p_i, ptr)$ of $P$}{
%     \If{$ptr \in M$}{
%         skip\;
%     }
%     \Else{
%         $T \leftarrow ptr$ \tcc*{Add to list}
%         $M \leftarrow ptr$ \tcc*{Insert in hashmap}
%     }
% }
% \Return{$T$}
% \end{algorithm}
%With this implementation of the search method, 
Thus, the returned tuples are  ones with at least one entry predicate matching the query predicate. This has important implications.
%that are important to discuss. 
Let the entries of a tuple $T$ be $E^T_i$ ($i \in \{1, \cdots , n\}$). Based on the query predicate $q$, two cases exist.
\begin{align}
& \textit{Consistent}(T, q) \Leftrightarrow \exists i: \textit{Consistent}(E^T_i, q) 
 \label{eq:consistent_1}\\
& \textit{Consistent}(T, q) \Leftrightarrow \forall i: \textit{Consistent}(E^T_i, q) \label{eq:consistent_2}
\end{align}
\Cref{eq:consistent_1} is valid for all operators that require at least one entry to match the query predicate. An example for this case is the {\bf spatial overlaps} operator. On the other hand, \Cref{eq:consistent_2} corresponds to operators that need all entry predicates to match the query predicate, e.g., strictly left or right, contained by, and disjoint.
%As said previously, t
The search algorithm returns all tuples that have at least one entry matching the query predicate, i.e., is equivalent to Eq.~(\ref{eq:consistent_1}). 
These queries can be answered efficiently. To answer queries of the second type, it is still possible to use the index, but it will be less efficient. Indeed, if all entries need to match the query predicate, at least one will match. The index can thus be used to find all the tuples that have at least one matching entry. In a second refinement step, the actual operator can then be rechecked on the returned tuples to select only the ones that actually match the query predicate.

%It is worth noting that many queries corresponding to Eq.~(\ref{eq:consistent_2}) can already be answered efficiently with traditional GiST or SP-GiST indices. For example, the strictly left ($\ll$) spatial operator requires that all parts of the first geometry are left of the second geometry. In this case, even if the geometries are collections of many points or polygons, it is sufficient to know the extent of the collection as a single bounding box to know if one collection is strictly left of the other collection or not. The MGiST and MSP-GiST indices are thus not meant to completely replace the existing GiST and SP-GiST indices but rather complement them.

\subsection{Nearest Neighbour Search}
\label{section:knn_search}
The search method  in Section~\ref{section:search} scans the index for tuples qualifying a given predicate. Another type of query that can be answered using GiST and SP-GiST indices is a k-nearest neighbour search (KNN). This query returns the k closest objects/tuples to a given query value based on a user-specified distance measure. Efficient algorithms exist to answer KNN queries efficiently using GiST and SP-GiST indices, e.g.,~\cite{knn_1,knn_2}. Similarly, existing algorithms can be adapted to answer KNN queries for MGiST and MSP-GiST.

To answer KNN queries with MGiST or MSP-GiST indices, the user specifies a distance measure to be applied between an index entry and a query q. Using this distance measure, the index finds the k-closest entries to q, corresponding to the k closest tuples. However, in multi-entry indices, 
%the search method will similarly find the closest entries to the query value. I
it is possible that multiple of these closest entries point to the same tuple. Thus, as in Section~\ref{section:search}, a de-duplication step is  necessary. De-duplication   retains only the closest entry for each distinct tuple. Thus, a KNN search in  MGiST or MSP-GiST works  only  for distance measures that satisfy \cref{eq:distance}. 
%In case global information of the tuple is required to compute the distance to a query value, the traditional GiST or SP-GiST indices can be used.
\begin{equation}
\textit{dist}(T, q) = \min_{i \in \{1, ..., n\}} \textit{dist}(E^T_i, q) \label{eq:distance}
\end{equation}
% The KNN search algorithm is omitted as it is identical to the traditional KNN search in GiST and SP-GiST with the addition of the same de-duplication step as in Algorithm~\ref{alg:search}.

\subsection{MGiST and MSP-GiST vs. GIN}
\label{section:gin}
In this section, we discuss the similarities and differences 
between MGiST and MSP-GiST in one side and another generalized index structure: GIN~\cite{gin} on the other side.
We further explain the use cases of each. GIN is a generalized inverted index that can be used to index arrays and documents~\cite{gin}. Its implementation also requires an {\bf ExtractValue} method that also splits a tuple into multiple sub-elements, termed keys (e.g., values in an array, or words in a document). 
%The name \textit{ExtractValue} is borrowed from GIN. 
This is the main similarity.
%: in both indices, the tuple gets split into multiple keys/entries before being indexed.
However, two main structural differences exist between MGiST/MSP-GiST and GIN indices. (1)~GIN indexes the keys in a B-Tree. This makes GIN not suitable for indexing complex spatial and spatio-temporal objects. (2)~The data indexed by GIN is assumed to contain many duplicate keys. For example, a word can be present in many different documents. To handle this, GIN indexes the distinct keys in a B-Tree and stores for each key at the leaf level a set of tuple pointers that contain this key. In MGiST and MSP-GiST, 
%this assumption is not made, and 
%the index values are thus key-pointer pairs ($E = (p, ptr)$). In case entries of different tuples have identical keys, they will still be 
entries with the same keys are stored as separate entries in the index leaves. 
%However, when working with spatial or spatio-temporal data, such as discussed in Section~\ref{section:traj_index}, this will almost never happen. 
Adding this de-duplication of keys/predicates to MGiST and MSP-GiST 
%could be useful, for example, when indexing document data in a multi-entry trie (which requires MSP-GiST). This 
is left as future work.
\section{Multi-Entry Trajectory Indexing}
\label{section:traj_index}
In this section, we illustrate the use of MGiST and MSP-GiST for indexing moving object data (also termed trajectory data) that can span over large spatio-temporal extents. 
Thus, indexing trajectories in a single bounding box in GiST or SP-GiST produces many false positives during query processing.
%For example, the extent of the trajectory of a car travelling from Amsterdam to Madrid would cover the whole of France and even London, which is not on the same landmass. The temporal extent of such a trip would also last almost a full day, if not more if the trip contains temporal gaps. The current GiST or SP-GiST indices index this trip as a single bounding box spanning the whole spatio-temporal extent of the trajectory. A query searching for all the cars passing through London between 10 am and 2 pm on the day of this trip will thus return the trajectory of the trip as a potential result. 
However, if each trajectory is indexed using a combination of multiple smaller bounding boxes, this problem can be mitigated. Refer to Figure~\ref{fig:intro}, where  a trajectory circles the city of Brussels. From this example, it is clear that trajectory indexing is a perfect use case for MGiST and MSP-GiST.
Thus, we instantiate multi-entry implementations for the R-Tree using MGIST, and a Quad-Tree, and KD-Tree indices using MSP-GiST. As in Section~\ref{section:mgist}, these indices require a user-defined ExtractValue method that splits the tuples before indexing. Given a trajectory, the goal of these splitting algorithms is to produce a set of small bounding boxes that (accurately) represent the shape of the trajectory. Note that ExtractValue receives a single trajectory at a time. It does not have global information about the other trajectories.

Multiple models exist for representing trajectory data. In this paper, we assume the model used in MobilityDB, an open-source moving objects database presented in~\cite{mobilitydb}. A trajectory is stored in a data type termed \textit{tgeompoint} that can have four temporal durations: instant, instant set, sequence, or sequence set. A tgeompoint instant represents a spatial position (Point) at a specific instant in time. A tgeompoint instant set groups multiple instants with distinct timestamps, but assumes no interpolation between instants. A tgeompoint sequence is similar to an instant set but assumes linear interpolation between subsequent instants. This is the data type that is used to represent continuous trajectories. Finally, a tgeompoint sequence set groups non-overlapping tgeompoint sequences and can be used, e.g., to represent a trajectory containing stops or signal losses.
We focus on the problem of splitting tgeompoint sequence data as it is the main data type  for trajectory data. 
%In case trajectories contain temporal gaps and are stored as sequence sets, the presented algorithms can be applied to the individual sequences. More sophisticated splitting algorithms for sequence sets are left as future work.
A trajectory is a tgeompoint sequence $T$, containing $n$ instants defined between $t_0$ and $t_{n-1}$. An instant is a position-timestamp pair ($p@t$), with the position being either 2D or 3D. A trajectory segment is defined as two consecutive instants of the trajectory. Thus, a  trajectory with $n$ instants is made of $n-1$ consecutive segments.
\begin{equation}
    T = [p_0@t_0, ..., p_{n-1}@t_{n-1}]
\end{equation}
At time $t$ between $t_0$ and $t_{n-1}$, Position $p(t)$ is computed using linear interpolation of Segment $[p_i@t_i, p_{i+1}@t_{i+1}]$ containing $t$ by:
\begin{equation}
    p(t) = \frac{p_i \cdot (t_{i+1} - t) + p_{i+1} \cdot (t - t_i)}{t_{i+1} - t_i}, t_i \le t < t_{i+1} \label{eq:traj_pos}
\end{equation}
There are many ways to split  a trajectory into multiple bounding boxes. These methods vary in the way the splits are generated and in how many boxes the trajectory is split. One straightforward splitting algorithm is the following. Given a trajectory $r$ and a constant parameter $k$ (the same for all tuples), split $r$ into $k$ `equal' parts. This method produces a box for every $\lceil \frac{n}{k} \rceil$ consecutive segments. 
%(The last box might contain less than $\lceil \frac{n}{k} \rceil$ segments.) 
Refer to this splitting algorithm by \textbf{EquiSplit}.
Notice that  {\bf EquiSplit } splits all trajectories into the same number of $k$ boxes. Given a dataset of $N$ trajectories, {\bf EquiSplit} produces in total $N \cdot k$ boxes, i.e., the size of the created index is directly proportional to $k$.

Another method to split trajectories into $k$ boxes is by Hadjieleftheriou et al.~\cite{mergesplit}, where it minimizes the total volume of the generated bounding boxes. Then, they propose an approximate version of the algorithm, termed \textbf{MergeSplit}, that produces near-optimal splits in $\mathcal{O}(nlogn)$ time ($n$ being the number of instants in the trajectory). 
%This algorithm is called \textbf{MergeSplit}. 
It starts by computing the bounding box of each individual trajectory segment. Then, it iteratively merges the two adjacent bounding boxes that produce the least increase in volume until the desired number of boxes is reached.

In both algorithms, the number of bounding boxes per trajectory is constant. This is a big restriction and 
%it is easy to imagine that this might not be 
is not 
optimal. In fact, in case some trajectories are much longer than others (have a bigger spatial extent), it could be beneficial to split them into more boxes. Thus, we propose an algorithm, termed \textbf{SegSplit}, that requires a parameter $m$; the number of segments that form a bounding box. A trajectory with $n$ segments is  split into $\lceil \frac{n}{m} \rceil$ boxes. \textbf{SegSplit} is  the inverse of {\bf EquiSplit}, which has a fixed number of boxes per trajectory but a variable number of segments per box.

Analogously,  {\bf MergeSplit}  can be adapted to generate a different number of boxes per trajectory. Given a parameter $m$ (same for all trajectories) and a trajectory with $n$ segments, we compute the desired number of boxes $k = \lceil \frac{n}{m} \rceil$. Then, we apply {\bf MergeSplit} given $k$. 
%Thus, this algorithm is to MergeSplit what ManualSplit is to EquiSplit. Let us call it 
We refer to this adaptive splitting algorithm by 
\textbf{AdaptSplit}.

Rasetic et al.~\cite{linearsplit} present a cost model for splitting trajectories based on the expected number of I/O's of a given range query. They  present an optimal splitting algorithm that minimizes the cost of a given range query. To avoid  quadratic time complexity, they propose a heuristic linear-time algorithm, termed \textbf{LinearSplit}, that produces near-optimal results. \textbf{LinearSplit} starts by linearly accumulating the trajectory instants until a given criterion is met. A constant-slope approximation of the collected sub-sequence is  used to compute the bounding boxes covering this part of the trajectory. This step takes linear time.
%, as computing the optimal split for a constant-slope trajectory can be solved analytically. 
\textbf{LinearSplit} continues with the remaining trajectory instants. For more detail, the readers are referred to~\cite{linearsplit}. Similarly to SegSplit and AdaptSplit, \textbf{LinearSplit} produces a variable number of boxes per trajectory.

\begin{figure}
    \centering
    \begin{subfigure}{.48\linewidth}
        \centering
        \includegraphics[width=\linewidth]{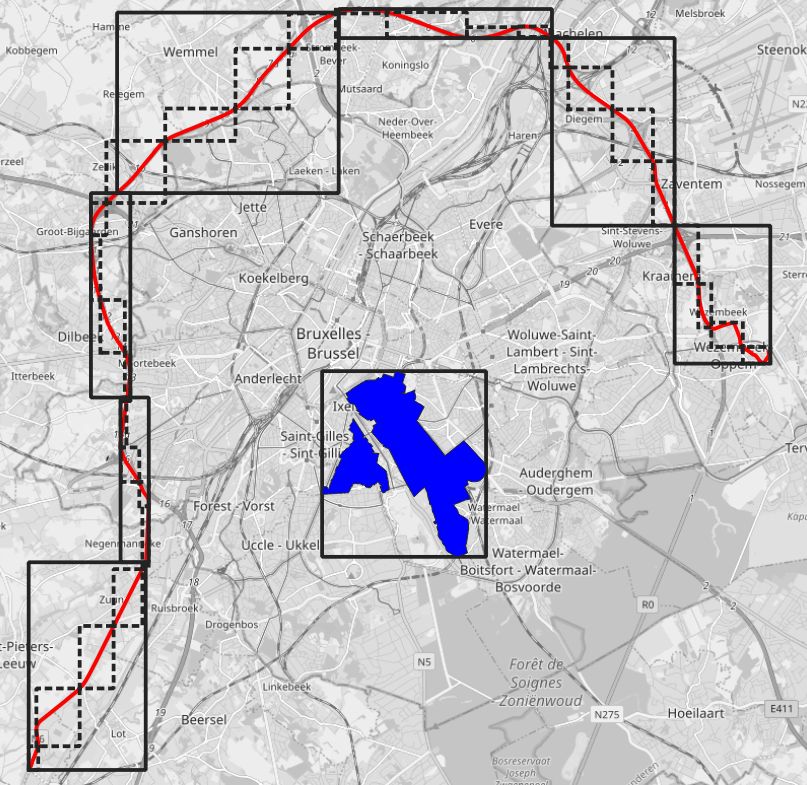}
        \caption{EquiSplit / SegSplit}
        \label{fig:equisplit}
    \end{subfigure}
    \begin{subfigure}{.48\linewidth}
        \centering
        \includegraphics[width=\linewidth]{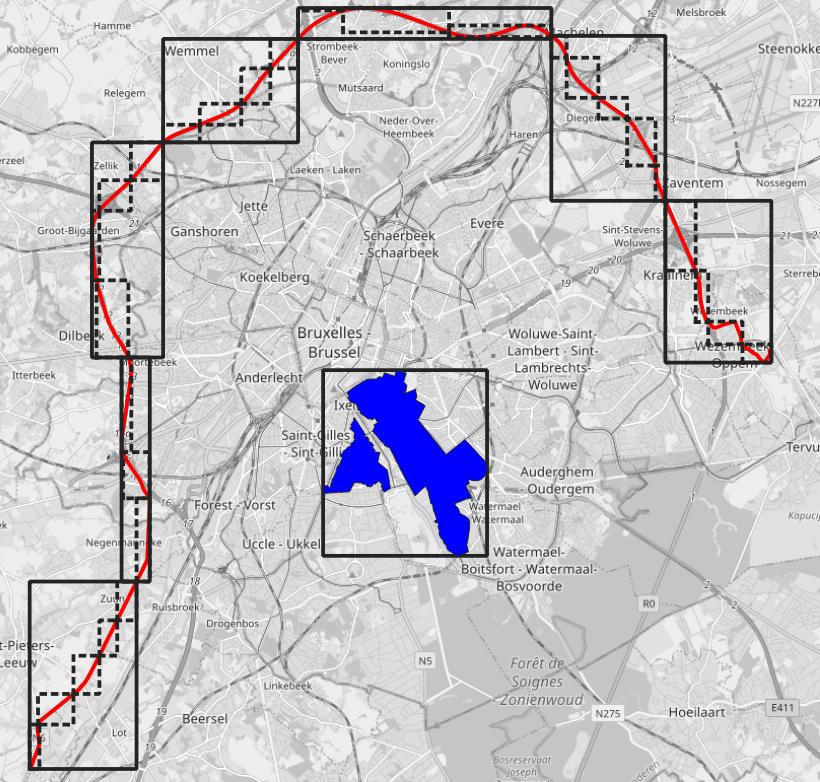}
        \caption{MergeSplit / AdaptSplit}
        \label{fig:mergesplit}
    \end{subfigure}
    \begin{subfigure}{.48\linewidth}
        \centering
        \includegraphics[width=\linewidth]{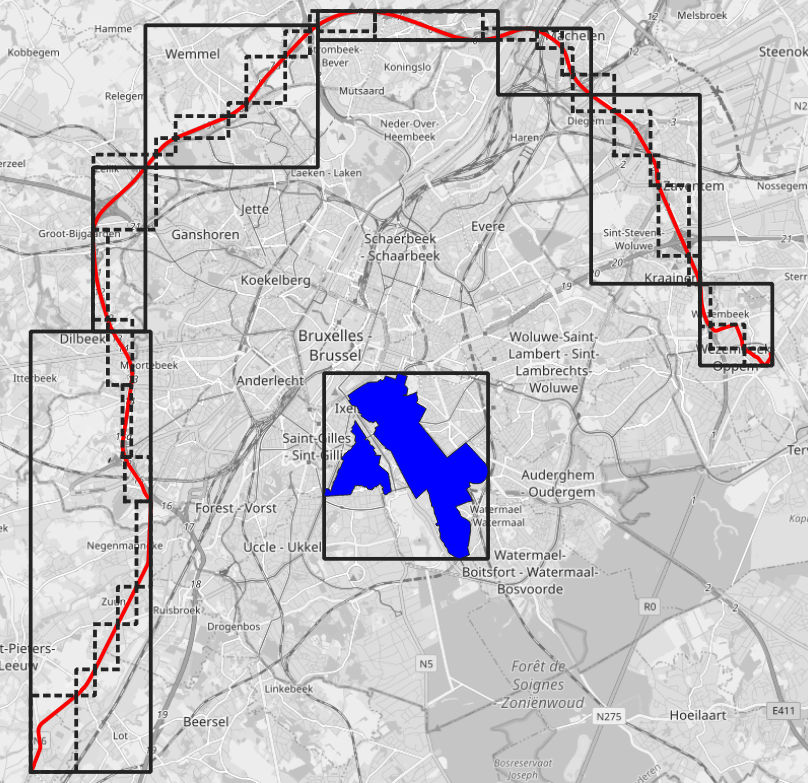}
        \caption{LinearSplit}
        \label{fig:linearsplit}
    \end{subfigure}
    \caption{Visualization of the bounding boxes resulting from using different splitting algorithms on a trajectory.}
    \label{fig:splitting_viz}
\end{figure}

Figure~\ref{fig:splitting_viz} illustrates the result of applying the different splitting algorithms on a single trajectory. Although the figure only illustrates the spatial projection of the bounding boxes, it's important to keep in mind that all splitting algorithms also consider the temporal dimension. Each image contains two splits, one in 7 bounding boxes (with solid border) and one in 28 boxes (dotted border). Note that for a single trajectory, EquiSplit and SegSplit are equivalent. Indeed, given a number of boxes $k$ and a trajectory $T$ of length $n$, one can compute the parameter $m = \lceil \frac{n}{k} \rceil$, such that EquiSplit($T$, $k$) and SegSplit($T$, $m$) produce the same boxes. The same holds for MergeSplit and AdaptSplit. However, this is not true anymore when applying the algorithm with the same parameter on multiple trajectories of different lengths.

All the splitting algorithms presented in this section assume that the splitting can only occur at instants stored in the sequence. In theory, if two subsequent instants are far apart, the line formed by linear interpolation of these two instants could also be split to further reduce the size of the created bounding boxes. Taking this into account, however, would greatly complicate the splitting algorithms. In practice, such a situation is also very rare when working with moving objects data. This is thus left as future work.

The splitting algorithms are used as implementations of the ExtractValue method to produce four variants of the multi-entry R-Tree, Quad-Tree, and KD-Tree implementations. These variants are evaluated in Section~\ref{section:experiments}. The remaining user methods (Consistent, Picksplit, etc.) needed to implement the multi-entry indices are all identical to their traditional counterpart. The only exception is the {\bf Compress} method. In  GiST and SP-GiST,  {\bf Compress}   receives a geometry or trajectory and produces a bounding box as a compressed version of the data. In MGiST and MSP-GiST,  {\bf Compress}  applies to all entries returned by ExtractValue. Since these methods already return a list of bounding boxes,  {\bf Compress}  is the identity.
\section{Experiments}
\label{section:experiments}

In this section, we evaluate the performance of the proposed multi-entry MGiST and MSP-GiST indices and their different splitting algorithms. In the first experiment, the four splitting algorithms are evaluated using the BerlinMOD benchmark for moving objects. We omit the evaluation of Equisplit, as Equisplit is  consistently outperformed by  the other four algorithms. We evaluate index construction in terms of both construction time and index size. Query speed of  point, range and KNN queries is reported and discussed. In a second experiment, the performance of the best index found is further evaluated on a real-world data set of AIS vessel trajectories.

\begin{figure}
    \centering
    \begin{subfigure}{.46\linewidth}
        \centering
        \includegraphics[width=\linewidth]{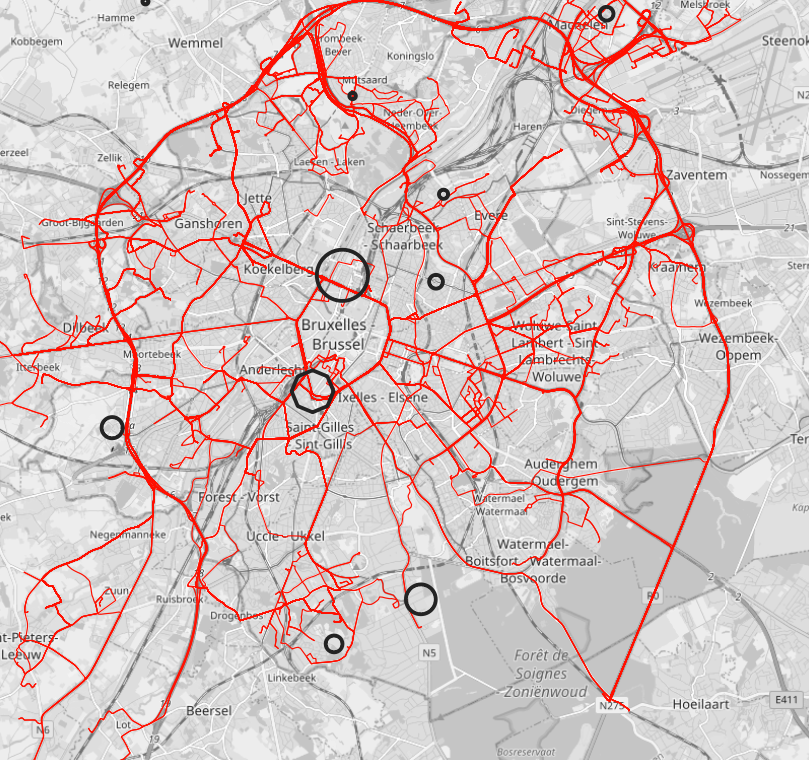}
        \caption{BerlinMOD}
        \label{fig:dataset_1}
    \end{subfigure}
    \begin{subfigure}{.48\linewidth}
        \centering
        \includegraphics[width=\linewidth]{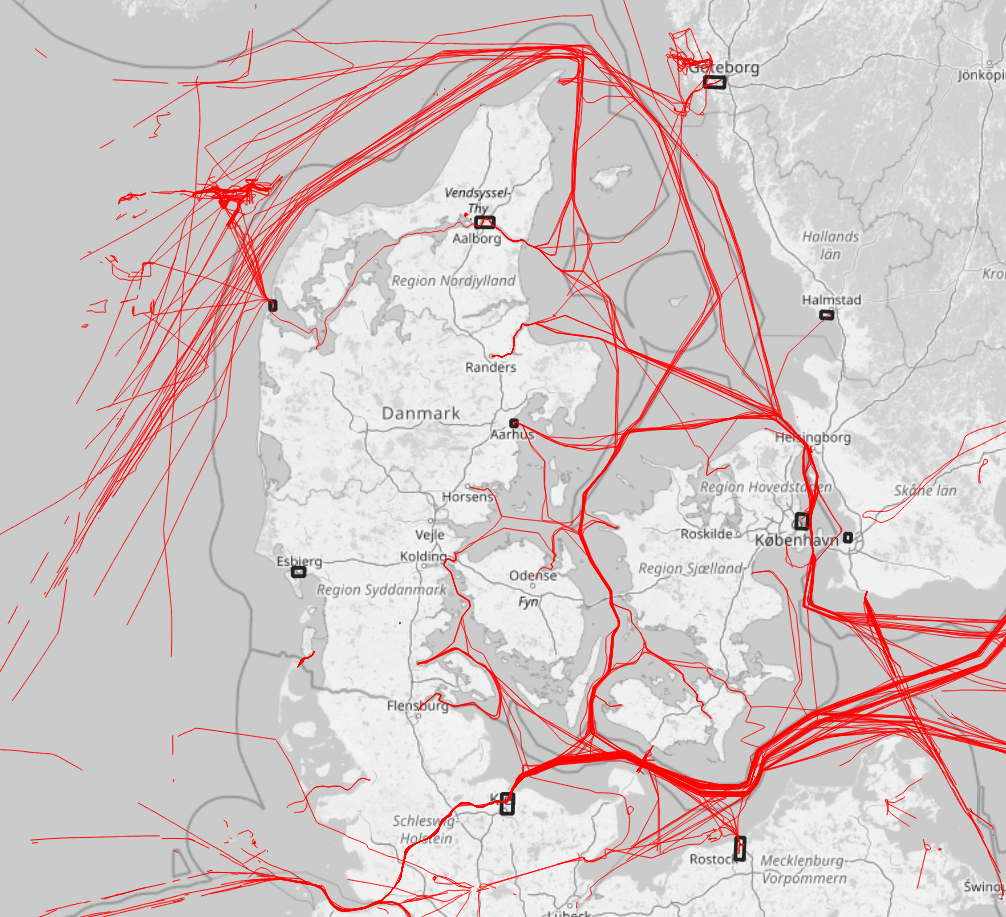}
        \caption{AIS}
        \label{fig:dataset_2}
    \end{subfigure}
    \caption{Data sets used in the experiments}
    \label{fig:dataset}
\end{figure}

\subsection{The BerlinMOD Benchmark}
\label{section:berlinmod}

BerlinMOD~\cite{berlinmod} is a benchmark for moving objects databases that presents both a synthetic trajectory data generator and a set of benchmark queries. In these experiments, we use a version of the BerlinMOD data generator that is implemented in MobilityDB, and thus generates the trajectories readily in MobilityDB moving point format.\footnote{\url{https://github.com/MobilityDB/MobilityDB-BerlinMOD}} The generator simulates work and leisure trips in and around the city of Brussels. It is parameterized by a scale factor that determines the size of the generated data set. We use a scale factor of 1 that produces trips for 2000 vehicles over 30 days, for a total of 157K trips containing on average 1.2K instants each. The size of this trajectory data set is 7.7GB. Next to the trajectory data, the generator produces four other tables of interest: \verb|QueryInstants|, \verb|QueryPeriods|, \verb|QueryPoints| and \verb|QueryRegions|, that are used in the benchmark queries. The query regions (Fig.~\ref{fig:dataset_1}, black) have an average size of 1 km$^2$. Table~\ref{tab:berlinmod_tables} summarizes the generated tables.

\begin{table}
    \caption{Tables produced by the BerlinMOD generator.}
    \begin{tabular}{lll}
    \toprule
    Name & Columns & Count \\
    \midrule
    Trips & $\langle$vehicle\_id, trip\_id, trip$\rangle$ & 157549 \\
    QueryInstants & $\langle$id, instant$\rangle$ & 100 \\
    QueryPeriods & $\langle$id, period$\rangle$ & 100 \\ 
    QueryPoints & $\langle$id, point$\rangle$ & 100 \\
    QueryRegions & $\langle$id, region$\rangle$ & 100 \\
    \bottomrule
    \end{tabular}
    \label{tab:berlinmod_tables}
\end{table}

\begin{figure*}
    \centering
    \includegraphics[width=\textwidth]{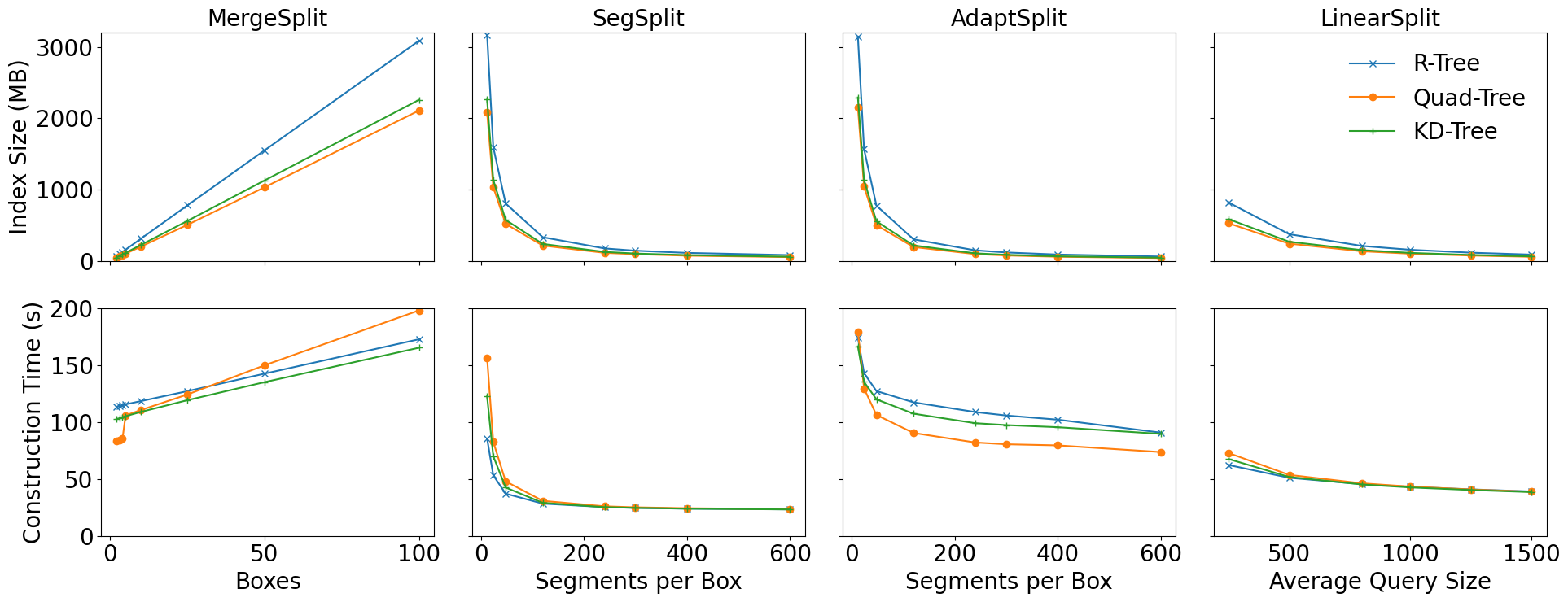}
    \caption{Index size and construction time of the different splitting algorithm.}
    \label{fig:construction}
\end{figure*}

A spatio-temporal index is constructed on the \verb|trip| column of the \verb|Trips| table, for different parameter values of the splitting algorithms. The parameter for MergeSplit is the number of boxes per trajectory. For SegSplit and AdaptSplit, the parameter is the (average) number of segments per box. Lastly, the parameter for LinearSplit corresponds to the average width, length and duration, in meters and minutes respectively, of the query box.

Figure~\ref{fig:construction} gives the index size and construction time for the four splitting algorithms. The experiments confirm linear proportionality between the number of splits and the index size in the case of MergeSplit, as mentioned in Section~\ref{section:traj_index}, and negative exponential in the case of SegSplit and AdaptSplit. For LinearSplit, a decrease in average query size results in an increase in number of generated boxes and thus an increase in index size.

Similarly, the index construction times are proportional to the index size (and thus to the number of stored boxes), with MergeSplit and AdaptSplit being relatively slower due to the $O(nlogn)$ complexity of the MergeSplit algorithm.

\subsubsection{Point and Range Queries}

The BerlinMOD benchmark has a total of 17 point and range query types. However, not all 17 query types make use of a spatio-temporal index on the \verb|Trips| table. We evaluate the following three point and range queries:

\noindent \textbf{Query 4}: Find the vehicles that passed by any of the query points.
\begin{verbatim}
SELECT DISTINCT vehicle_id, trip_id 
FROM Trips, QueryPoints 
WHERE eintersects(trip, point) 
AND trip && stbox(point);
\end{verbatim}

\noindent \textbf{Query 11}: Find the vehicles that passed by any of the query points at one of the query instants.
\begin{verbatim}
SELECT DISTINCT vehicle_id, trip_id 
FROM Trips, QueryPoints10, QueryInstants10 
WHERE eintersects(atTime(trip, instant), point)
AND trip && stbox(point, instant);
\end{verbatim}

\noindent \textbf{Query 13}: Find all the vehicles that passed through any of the query regions during one of the query periods.
\begin{verbatim}
SELECT DISTINCT vehicle_id, trip_id
FROM Trips, QueryRegions10, QueryPeriods10
WHERE eintersects(atTime(trip, period), region))
AND trip && stbox(region, period);
\end{verbatim}

In the above queries, the \verb|eintersects| predicate determines if a temporal point ever intersects a given geometric object. The function \verb|atTime| is used to restrict a temporal point to a given instant or period. Table \verb|QueryPoints10| contains only the first 10 rows of \verb|QueryPoints| and the same holds for the three other \verb|Query*10| tables. These tables are part of the BerlinMOD benchmark, and are used to limit the size of the Cartesian product when multiple \verb|Query*| tables are being joined. The last condition of every query tests the overlap between the trips and a spatio-temporal box (spatial only in the case of Query 4). This condition triggers the query optimizer to use the spatio-temporal index created on the trips.

\begin{figure*}
    \centering
    \includegraphics[width=\textwidth]{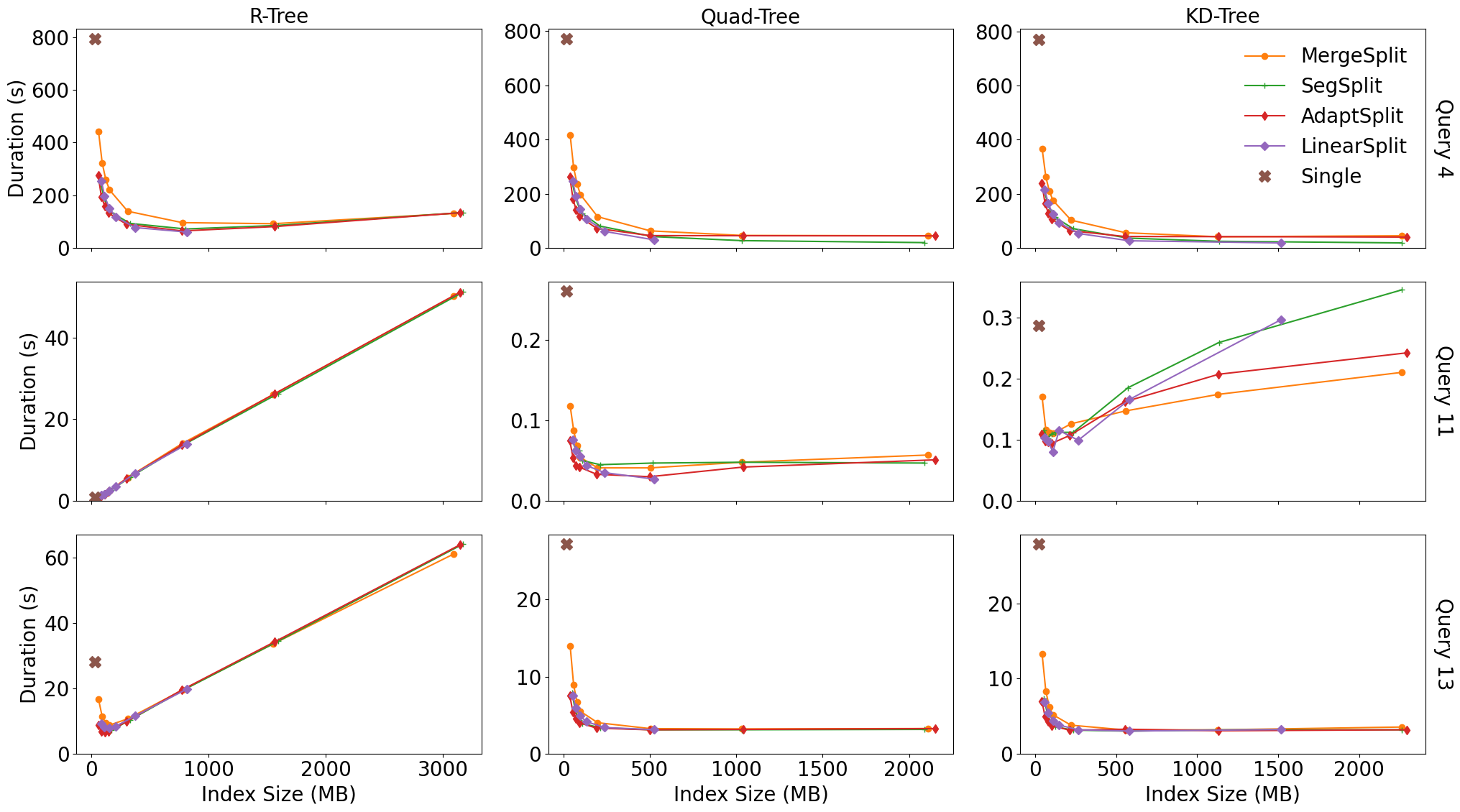}
    \caption{Point and range query duration for the multi-entry R-Tree, Quad-Tree and KD-Tree.}
    \label{fig:queries}
\end{figure*}

The cost of a point or range query utilising an index combines the cost of the index scan and the cost of applying the actual predicate on the tuples returned by the index, i.e., the refine cost. The refine cost is also determined by two factors: The cost of the functions and operators in the query, and the number of tuples returned by the index.
The main purpose of a multi-entry index is to reduce the number of tuples returned by the index, i.e., to improve the index filtering. However, this is at the cost of larger indices, and thus higher index scan cost. Based on this analysis, we can expect that the indices will behave better for queries with expensive functions and operators. Indeed, this is the case when reducing the number of tuples returned by the index becomes most beneficial.

Figure~\ref{fig:queries} gives the query duration of all tested indices for the three queries. Since the splitting algorithms are parameterized differently, we use the index size in the x-axis. The corresponding single-entry index is also shown.

The above analysis is coherent with the presented results. The most expensive query in terms of functions is Query 4 that applies an intersection test on the complete trajectories returned by the index. This index is also the one showing the most speedup when using multi-entry indices. 

The least expensive query is Query 11. This query applies a point intersection test at a snapshot of the returned trajectories. For this query, the multi-entry R-Tree implementations is actually slower than its traditional single-entry counterparts. In this case, the index scan is dominating the query duration, and is thus slowing down the query. The same effect can be seen for the KD-Tree at large index sizes, but to a lesser extent. The multi-entry MSP-GiST indices thus seem to have a lower index scan cost, as they are still able to improve the duration of Query 11. The increased index scan cost of the R-Tree indices is potentially due to the sequential insert method used by MGiST producing R-Trees with high overlap between nodes.

\begin{figure*}
    \centering
    \includegraphics[width=\textwidth]{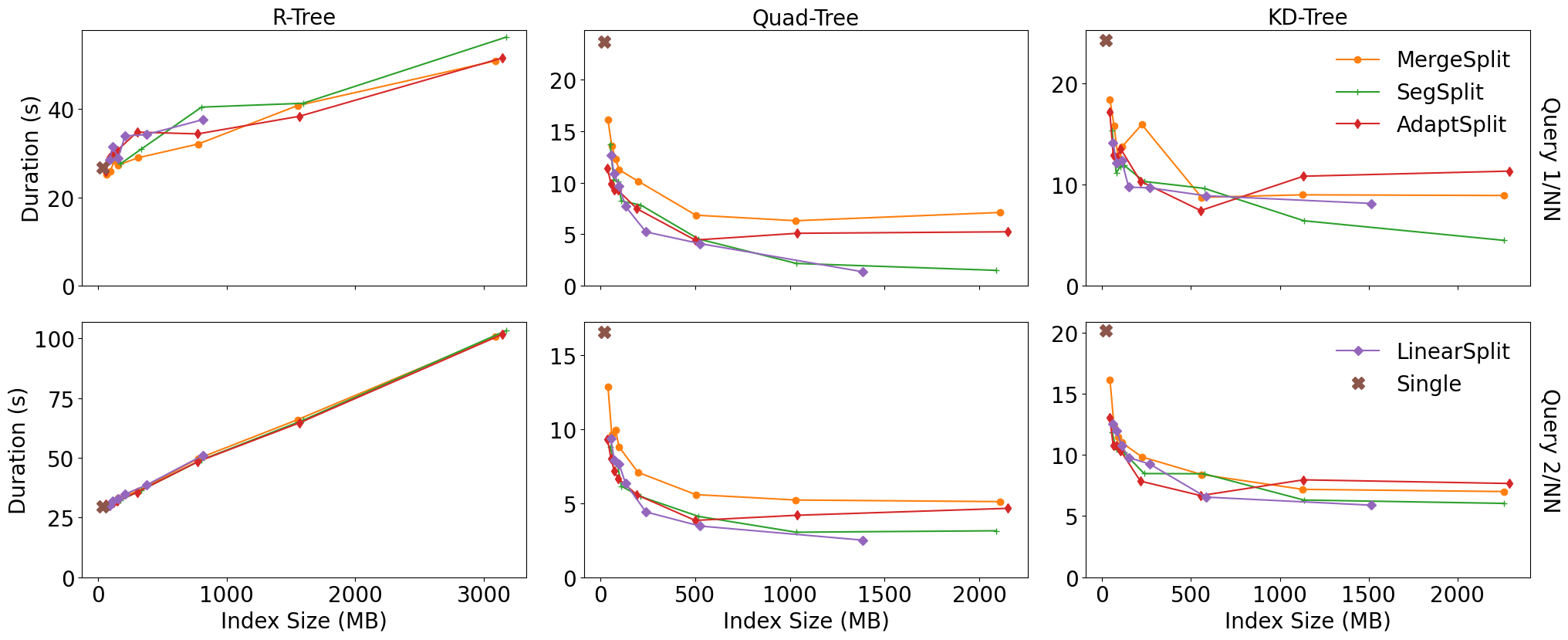}
    \caption{KNN query duration for the multi-entry R-Tree, Quad-Tree and KD-Tree.}
    \label{fig:knn_queries}
\end{figure*}

\subsubsection{Nearest-Neighbor Search Queries}

Nearest-neighbor search in trajectories can have multiple semantics. A historical continuous KNN query for moving objects is presented in~\cite{berlinmod}. This query searches for the k closest trajectories to a given spatio-temporal object at each instant that this object is defined. That is, the result of this query is also evolving in time and is returned as tuples of $(\textit{mobj}, t_s, t_e)$, where \textit{mobj} is the moving object and $(t_s, t_e)$ is the start and end times during which \textit{mobj} is part of the k closest trajectories. However, this continuous KNN query cannot be expressed using traditional SQL syntax, and is therefore not implemented in existing moving objects databases. Thus, we restrict the experiments to static KNN queries for moving objects. A static KNN query searches for the trajectories with the smallest \textit{nearestApproachDistance} to a given spatio-temporal object. We evaluate the following two KNN queries:

\noindent \textbf{Query 1/NN}: For each query point, find the 5 vehicles that passed closest to it.
\begin{verbatim}
SELECT point, vehicle_id, trip_id
FROM QueryPoints10
CROSS JOIN LATERAL (
    SELECT vehicle_id, trip_id FROM Trips 
    ORDER BY trip |=| stbox(point) 
    LIMIT 5) ClosestTrips;
\end{verbatim}

\noindent \textbf{Query 2/NN}: For each combination of query point and period, find the 5 vehicles that passed closest to the point during that period.
\begin{verbatim}
SELECT point, period, vehicle_id, trip_id
FROM QueryPoints10, QueryPeriods10
CROSS JOIN LATERAL (
    SELECT vehicle_id, trip_id FROM Trips 
    WHERE trip && stbox(period)
    ORDER BY trip |=| stbox(point, period) 
    LIMIT 5) ClosestTrips;
\end{verbatim}

These queries find the 5 trajectories with the closest point of approach to the query point (Query 1/NN) or the spatio-temporal points constructed by joining the query points and the query periods (Query 2/NN). The first query is essentially a spatial-only KNN query, while the second query is applied on temporal ranges. For clarity, we set $k$ to 5, though tests with $k$ up to two orders of magnitude larger yielded similar performance gains.

Figure~\ref{fig:knn_queries} gives the query duration of the two KNN queries for the evaluated indices. As can be seen, the R-Tree has poor query performance with increased index size, similarly to its performance in Query 11. Both the multi-entry Quad-Tree and KD-Tree showcase a large speedup over their single-entry counterparts, with the Quad-Tree performing the best overall. This is analogous to the results for the point and range queries.

\subsubsection{Comparison with Single-Entry Indices}

\begin{figure}
    \centering
    \includegraphics[width=\linewidth]{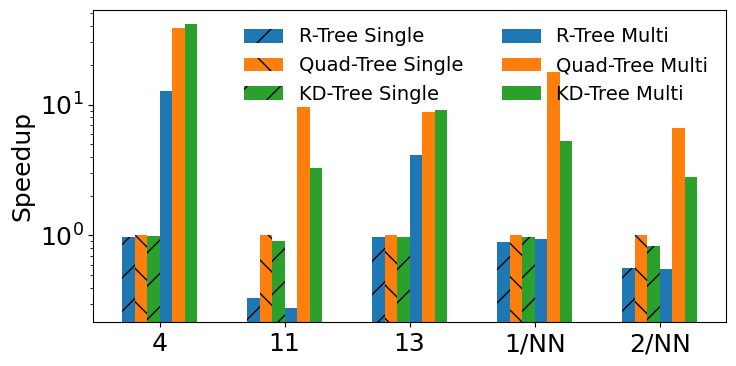}
    \caption{Speedup of the multi-entry indices compared to traditional single-entry indices. The best-performing single-entry index (Quad-Tree) is used as reference (speedup of 1).}
    \label{fig:speedup}
\end{figure}

For all five queries, the speedup obtained using the best-performing index is summarized in Figure~\ref{fig:speedup}. In this figure, the speed of each index is compared to the best-performing single-entry index, which for all queries is the single-entry Quad-Tree (Note the log scale). From this figure, the multi-entry Quad-Tree consistently provides equivalent or better query performance than the R-Tree and KD-Tree, with speedups up to $38\times$ for Query 4. Thus, the remaining experiments on AIS data are  applied on the multi-entry Quad-Tree only.

Concerning the splitting algorithms, MergeSplit is consistently performing the worst. This is due to the fact that it splits each trajectory into an equal number of bounding boxes. This  is not optimal when working with trajectories of various sizes. Secondly, the three remaining algorithms are approximately equivalent on Queries 4 and 13, with SegSplit performing slightly worse on Query 11. We further evaluate these three algorithms on real-world trajectory data in Section~\ref{section:ais}. 

\subsection{Vessel Trajectory Data}
\label{section:ais}

In this section, we further evaluate the performance of the multi-entry Quad-Tree on a real-world data set. The Danish Maritime Authority gives access to historical AIS data around Denmark.\footnote{\url{http://web.ais.dk/aisdata/}} In this experiment, we use the AIS data for the month of January 2023. The raw data set is 54.4GB. After importing the data into MobilityDB and applying basic data cleaning, the remaining data set contains 53K trajectories having on average 2.9K instants each. We also manually construct two additional tables: a table containing 10 ports around Denmark (Fig.~\ref{fig:dataset_2}, black) with and average size of 11.5 km$^2$, and a table containing the periods from 2pm to 4pm for Week 1 of January 2023. Table~\ref{tab:ais_tables} lists these tables.

\begin{table}
    \centering
    \caption{Tables used for the AIS experiment.}
    \begin{tabular}{lll}
    \toprule
    Name & Columns & Count \\
    \midrule
    Ships & $\langle$mmsi, trip\_id, trip$\rangle$ & 53008 \\
    Ports & $\langle$port, region$\rangle$ & 10 \\
    Periods & $\langle$day, period$\rangle$ & 7 \\
    \bottomrule
    \end{tabular}
    \label{tab:ais_tables}
\end{table}

We construct multi-entry Quad-Tree indices on the \verb|trip| column of the \verb|Ships| table, for the SegSplit, AdaptSplit and LinearSplit algorithms. We evaluate the following 2 range queries:

\noindent \textbf{Query 1}: Find the vessels that entered a port
\begin{verbatim}
SELECT mmsi, trip_id, port FROM Ships, Ports 
WHERE eintersects(trip, region) 
AND trip && stbox(region);
\end{verbatim}

\noindent \textbf{Query 2}: Find the vessels that entered a port between 2pm and 4pm on the first week of January 2023
\begin{verbatim}
SELECT mmsi, trip_id, port, day
FROM Ships, Ports, Periods
WHERE eintersects(atTime(trip, period), region) 
AND trip && stbox(region, period);
\end{verbatim}

The overlaps predicate triggers the optimizer to invoke the index. Query 1 contains a spatial-only intersection test, while Query 2 is fully spatio-temporal. These two queries are equivalent to Queries 4 and 13 of the BerlinMOD benchmark, respectively. 

\begin{figure}
    \centering
    \includegraphics[width=\linewidth]{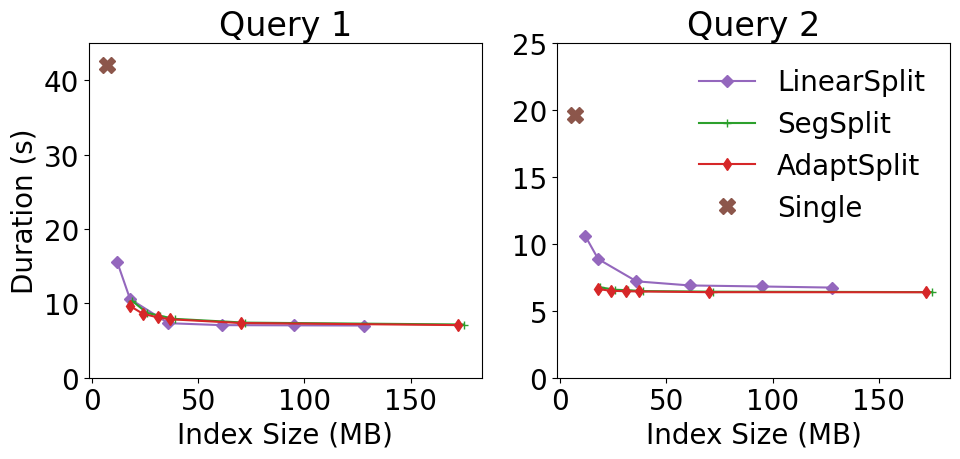}
    \caption{Duration of range queries on the AIS data set}
    \label{fig:ais_queries}
\end{figure}

The duration of both queries is given in Figure~\ref{fig:ais_queries}. The multi-entry Quad-Tree is able to provide a $6x$ and $3x$ speedup on Queries 1 and 2, respectively, with the three splitting algorithms producing similar results. Notice that in this experiment it is not beneficial to split trajectories into more than 5 boxes on average. In fact, past this point, the performance gain starts to plateau while the storage cost of the index continues to increase. A similar effect can also be seen in the BerlinMOD experiments. However, in this case, the cutoff is closer to 10 boxes per trajectory.

Thus, a good rule of thumb is  to construct a multi-entry Quad-Tree using one of the three best splitting algorithms, and using parameters such that the trajectories are split on average into 5 to 10 (or more) boxes. For MergeSplit and AdaptSplit, the parameter can be computed by dividing the average length of the trajectories $\Tilde{n}$ by the desired number, e.g., $k = \frac{\Tilde{n}}{10}$.

\section{Discussion} 
\label{section:discussion}

We introduce MGiST and MSP-GiST as a solution to construct efficient trajectory indices. However, this solution is not limited to trajectory data. Any type of composite data that can be decomposed into smaller parts can make use of MGiST and MSP-GiST. For example, checking for the intersection between a trajectory and a static point is almost identical to checking the intersection between a linestring formed by the spatial projection of the trajectory and this point. This is a spatial-only operation. Thus, MGiST and MSP-GiST can  provide benefits to spatial-only indices. 

A specific use-case of MGiST and MSP-GiST is to index PostGIS line strings, polygons, and geometry collections. A splitting algorithm for line strings would be similar to the ones discussed in Section~\ref{section:traj_index} but would have to take into account that line strings can have self-crossings. This is not the case in trajectories, as the time dimension is always increasing. Checking for overlap between complex polygonal shapes could be made more efficient by indexing the convex parts of the stored polygons. Lastly, if a database column stores geometry collections, each geometry can be indexed in a separate index entry. A typical example of this would be the indexing of island groups stored as multi-polygons.

Another data type that can benefit from multi-entry indexes is time-series data. Time-series data can be viewed as 1D equivalent to trajectories. Previous work~\cite{timeseries} discusses ways to split time-series data into segments for better indexing. Using MGiST and MSP-GiST, these solutions can be implemented with little effort.

Section~\ref{section:gin} presents two main structural differences between GIN and MGiST/MSP-GiST. Another behavioural difference is the fact that  GIN also allows queries to split their query value into multiple query predicates before searching the index. This allows GIN to answer questions such as: Which documents contain this specific sentence? The sentence is  split into words before querying the GIN index. Splitting is done via a key method called \textbf{ExtractQuery}. This ExtractQuery method could also be useful for trajectory, geometry, or time-series data. For example, when looking for all trajectories crossing the border between two countries, the query value is a line string. Splitting this line string in bounding boxes before querying the trajectory index could further reduce false positives and decrease query times. We leave the implementation of the ExtractQuery method as future work.

\section{Conclusion}
\label{section:conclusion}

This paper introduces two generalizable index structures MGiST and MSP-GiST. They allow the indexing of complex or composite multidimensional data types by decomposing them into smaller parts before being indexed in a traditional GiST or SP-GiST index. This splitting mechanism is exposed to the user as a pluggable module, similar to the existing GiST and SP-GiST modules. This allows the index to be tailored to specific data types and operators.

We then detailed how the MGiST and MSP-GiST index can be used to efficiently index trajectory data as one example of complex multidimensional data. Using the pluggable modules, we implemented two existing and two newly proposed splitting mechanisms to build multi-entry variants of three well-known indices, the R-Tree, Quad-Tree and KD-Tree. Evaluations on synthetic and real-world data sets showcase significant speedups compared to the single-entry counterparts for point, range and KNN queries.

\begin{acks}
Walid G. Aref acknowledges the support of the National Science Foundation under Grant Number  IIS-1910216.
Maxime Schoemans is a Research Fellow of the Fonds de la Recherche Scientifique - FNRS.
\end{acks}

\bibliographystyle{ACM-Reference-Format}
\bibliography{references}

\end{document}